\documentclass[preprint,aps]{revtex4}
\usepackage{mathrsfs}
\usepackage{amsmath}
\usepackage{graphicx}
\usepackage{multirow}
\usepackage{makecell}
\usepackage{booktabs}

\begin{document}

\title{Spectroscopic Evidence of Bilayer Splitting and Interlayer Pairing in an Iron Based Superconductor}

 \author{ Dingsong Wu$^{1,2,\sharp}$, Wenshan Hong$^{1,2,\sharp}$, Chenxiao Dong$^{1,2,\sharp}$, Xianxin Wu$^{6,1}$, Qiangtao Sui$^{1,2}$, Jianwei Huang$^{1,2}$, Qiang Gao$^{1,2}$, Cong Li$^{1,2}$, Chunyao Song$^{1,2}$, Hailan Luo$^{1,2}$, Chaohui Yin$^{1,2}$, Yu Xu$^{1,2}$, Xiangyu Luo$^{1,2}$, Yongqing Cai$^{1,2}$, Junjie Jia$^{1,2}$, Qingyan Wang$^{1}$, Yuan Huang$^{1}$, Guodong Liu$^{1,4}$, Shenjin Zhang$^{3}$, Fengfeng Zhang$^{3}$, Feng Yang$^{3}$, Zhimin Wang$^{3}$, Qinjun Peng$^{3}$, Zuyan Xu$^{3}$, Xianggang Qiu$^{1,2}$, Shiliang Li$^{1,2}$, Huiqian Luo$^{1,4*}$, Jiangping Hu$^{1,2,4*}$, Lin Zhao$^{1,4*}$ and X. J. Zhou$^{1,2,4,5,*}$}

\affiliation{
\\$^{1}$Beijing National Laboratory for Condensed Matter Physics, Institute of Physics, Chinese Academy of Sciences, Beijing 100190, China.
\\$^{2}$University of Chinese Academy of Sciences, Beijing 100049, China.
\\$^{3}$Technical Institute of Physics and Chemistry, Chinese Academy of Sciences, Beijing 100190, China.
\\$^{4}$Songshan Lake Materials Laboratory, Dongguan, Guangdong 523808, China.
\\$^{5}$Beijing Academy of Quantum Information Sciences, Beijing 100193, China.
\\$^{6}$Department of Physics, the Pennsylvania State University, University Park, PA, 16802, USA.
}

\date{January 13, 2020}

\maketitle

{\bf
In high temperature cuprate superconductors, the interlayer coupling between the CuO$_2$ planes plays an important role in dictating superconductivity, as indicated by the sensitive dependence of the critical temperature (T$_C$) on the number of CuO$_2$ planes in one structural unit\cite{MGreven2004HEisaki}. In Bi$_2$Sr$_2$CaCu$_2$O$_{8+\delta}$ superconductor with two CuO$_2$ planes in one structural unit, the interaction between the two CuO$_2$ planes gives rise to band splitting into two Fermi surface sheets (bilayer splitting)\cite{ZXShen2001PVBogdanov,DLFeng2001ZXShen,DSDessau2001YDChuang} that have distinct superconducting gap\cite{XJZhou2019PAi}. The iron based superconductors are composed of stacking of the FeAs/FeSe layers; whether the interlayer coupling can cause similar band splitting and its effect on superconductivity remain unclear. Here we report high resolution laser-based angle-resolved photoemission spectroscopy (ARPES) measurements on a newly discovered iron based superconductor, KCa$_2$Fe$_4$As$_4$F$_2$ (T$_C$=33.5\,K) which consists of stacking
FeAs blocks with two FeAs layers separated by insulating Ca$_2$F$_2$ blocks. Bilayer splitting effect is observed for the first time that gives rise to totally five hole-like Fermi surface sheets around the Brilliouin zone center. Band structure calculations reproduce the observed bilayer splitting by identifying interlayer interorbital interaction between the two FeAs layers within one FeAs block. All the hole-like pockets around the zone center exhibit Fermi surface-dependent and nodeless superconducting gap. The gap functions with short-range antiferromagetic fluctuations are proposed and the gap symmetry can be well understood when the interlayer pairing is considered. The particularly strong interlayer pairing is observed for one of the bands. Our observations provide key information on the interlayer coupling and interlayer pairing in understanding superconductivity in iron based superconductors.

 }

\vspace{3mm}

     The newly discovered superconductor, ACa$_2$Fe$_4$As$_4$F$_2$ (A=K, Rb and Cs, 12442), has attracted much attention because it provides a unique platform to study superconductivity mechanism of the iron based superconductors\cite{GHCao2016ZCWang,GHCao2017ZCWang,XBShi2016GTWang,ADHillier2017JIshida,SJBlundell2018FKKKirschner,SJBlundell2018MSmidman,CBernhard2019BXu,SYLi2019YYHuang,GMu2019TWang,GHCao2019ZCWang}. On the one hand, its crystal structure consists of bilayer FeAs blocks separated by insulating Ca$_2$F$_2$ blocks\cite{GHCao2016ZCWang,GHCao2017ZCWang} (Fig. 1a). The iron based superconductors discovered so far can be classified into three main groups: one group consisting of single FeAs layer separated by insulating block layers (left panel in Fig. 1b) such as LaFeAsO system\cite{HHosono2008YKamihara}; another group consisting of infinite stack of FeAs/FeSe layers (right panel in Fig. 1b) like FeSe\cite{MWu2008FHsu}, LiFeAs\cite{CQJin2008XCWang} and BaFe$_2$As$_2$\cite{DJohrendt2008MRotter} systems. 12442 represents the third group that contains bilayer FeAs blocks separated by insulating Ca$_2$F$_2$ blocks (middle panel in Fig. 1b). This is analogous to the double CuO$_2$ layers separated by insulating blocks in Bi$_2$Sr$_2$CaCu$_2$O$_{8+\delta}$ (Bi2212) superconductor\cite{RBernard2013MClaude}. In Bi2212, the interlayer interaction between the two equivalent CuO$_2$ planes in one structural unit leads to band splitting and the formation of two Fermi surface sheets (bonding and antibonding)\cite{ZXShen2001PVBogdanov,DLFeng2001ZXShen,DSDessau2001YDChuang}. 12442 provides an opportunity to investigate whether the interaction between the two FeAs layers within one bilayer FeAs block can produce a similar bilayer splitting and the effect of interlayer coupling on superconductivity in iron based superconductors. On the other hand, band structure calculations indicate that there are six hole-like pockets at the centre of Brillouin zone (BZ) and four electron-like pockets at the zone corner\cite{XBShi2016GTWang,ADHillier2017JIshida}. The coexistence of such multiple Fermi surface sheets in one superconductor facilitates to study the relationship between the superconducting gap symmetry and the Fermi surface topology and to identify possible superconducting pairing model. The gap symmetry of 12442 reported so far has been controversial; muon-spin rotation experiment suggests it has a nodal gap\cite{SJBlundell2018FKKKirschner,SJBlundell2018MSmidman} while optical and transport experiments point to a nodeless gap\cite{CBernhard2019BXu,SYLi2019YYHuang,GMu2019TWang}. It becomes imperative to perform angle-resolve photoemission (ARPES) experiments on 12442 to directly measure its electronic structure and superconducting gap.

     In this paper, we report the first high-resolution angle-resolved photoemission measurements on the band structure and superconducting gap of KCa$_2$Fe$_4$As$_4$F$_2$ (K12442) superconductor. A complete Fermi surface topology is observed which consists of five hole-like pockets around the Brilliouin zone center $\Gamma$ point and tiny electron-like pockets around the zone corner M point. No Fermi surface nesting condition is satisfied between the electron-like pockets around M point and the hole-like pockets around $\Gamma$ point because of the apparent mismatch of their Fermi surface sizes. Band structure calculations indicate that these multiple bands around the zone center can be understood in terms of bilayer splitting where the interlayer interorbital interaction between the two FeAs layers within one bilayer FeAs block gives rise to band splitting. This is similar to the bilayer splitting observed in Bi2212 superconductor\cite{ZXShen2001PVBogdanov,DLFeng2001ZXShen,DSDessau2001YDChuang}. Superconducting gap is measured on all the observed Fermi surface sheets and Fermi surface-dependent and nodeless supercondcuting gap is revealed. We found that the superconducting gap is dramatically different on one set of the bilayer-split hole-like Fermi surface sheets; the maximal gap size shows up on one of the two sheets. The superconducting gap structure is understood by taking into account the interlayer pairing in the gap functions for the short-range antiferromagetic fluctuations. Our results provide key insights in the interlayer interaction and the interlayer pairing in iron based superconductors.

    High-quality single crystals of KCa$_2$Fe$_4$As$_4$F$_2$ were grown by the KAs flux method\cite{GMu2019TWang}. The samples were characterized by electrical resistivity (Fig. 1c) and magnetic susceptibility (Fig. 1d) measurements and the measured T$_C$ is 33.5\,K with a narrow transition width of $\sim$1\,K. High-resolution angle-resolved photoemission measurements were carried out on our two lab based ARPES systems. One is equipped with a hemispherical analyser DA30 (Scienta Omicron) and a helium lamp with a photon energy of h$\nu$ = 21.218\,eV (helium I)\cite{XJZhou2018}. The energy resolution was set at 10\,meV for the Fermi surface measurement and 2.5\,meV for the superconducting gap measurement. The other is equipped with a time-of-flight electron energy analyzer (ARToF 10k by Scienta Omicron) with a vacuum-ultra-violet (VUV) laser light source of h$\nu$= 6.994\,eV\cite{XJZhou2018,XJZhou2017YZhangNC}. This latest-generation ARToF-ARPES system is capable of measuring electronic structure covering two-dimensional momentum space (kx, ky) simultaneously. Another advantage of the ARToF analyser is that it has much weaker non-linearity effect so that the measured signal is intrinsic to the sample. The energy resolution was set at 1\,meV and the angular resolution was $\sim$0.3$^\circ$ corresponding to 0.004\,${\AA}^{-1}$ momentum resolution at the photon energy of 6.994\,eV. All the samples were cleaved \emph{in situ} at low temperature of 13\,K and measured in ultrahigh vacuum with a base pressure better than 5 $\times$ 10$^{-11}$\,mbar. The Fermi level was referenced by measuring on clean polycrystalline gold that was electrically connected to the sample or checked by the Fermi level of the measured sample in the normal state.

    Figure 1 shows the Fermi surface mapping of K12442 by using both laser (Fig. 1e to 1j) and helium lamp (Fig. 1k to 1m) light sources. The corresponding band structures are shown in Fig. 2. The helium lamp measurement can cover a large momentum space, giving an overall Fermi surface of K12442 in the normal state measured at 38\,K (Fig. 1m). Three hole-like pockets are clearly seen around $\Gamma$  point, labeled as $\alpha$, $\beta$ and $\gamma$ in Fig. 1m. Around M point, a tiny pocket ($\delta$) can be identified which is surrounded by four strong spots ($\epsilon $) (Fig. 1k and 1m). In the constant energy contour (CEC) at a binding energy of 20\,meV, the tiny pocket around M disappears and the four strong spots change into four circular pockets (Fig. 1l). Such an evolution indicates the tiny pocket around M is electron-like while the four strong spots are hole-like. In order to further examine on the Fermi surface topology around M, we carried out detailed momentum-dependent band structure measurements, as shown in Fig. 2b. Electron-like bands near M can be seen from the momentum Cut 3 to Cut 5 in Fig. 2b; their bottoms barely touch the Fermi level forming a tiny electron-like elliptical Fermi pocket as schematically shown in Fig. 1n (the other electron pocket is invisible due to matrix element effects and is shown as a dashed line). On the other hand, hole-like inverse-parabolic bands are observed for the momentum cuts (Cuts 2, 4 and 6) crossing the four strong spots. The top of these bands is close to but does not cross the Fermi level to form Fermi surface, as seen from Cut 2 and Cut 6 in Fig. 2b. The extended spectral weight of the band tops at the Fermi level gives rise to four strong spots around M as observed in Fig. 1k and 1m.

    Laser ARToF-ARPES experiments were carried out to measure the Fermi surface of K12442 around $\Gamma$ point.  The capability of covering two-dimensional momentum space simultaneously with much higher energy and momentum resolutions made it possible for us to reveal fine details on the electronic structure-band splitting in K12442. Two different laser polarizations were used in order to capture complete electronic structure by considering photoemmission matrix element effects\cite{Damascelli2003}. The measured Fermi surface mappings are shown in Fig. 1e-1j and the corresponding angle dependent band structures are shown in Fig. 2c-2h. All the three hole pockets are observed clearly with Fig. 1e-1h covering the inner ($\alpha$) and middle ($\beta$) pockets while Fig. 1i-1j covering the middle ($\beta$) and outer ($\gamma$) pockets. Moreover, band splitting is observed for the $\beta$ and $\gamma$ bands. The $\beta$ band splitting can be clearly seen from the constant energy contour in Fig. 1f where the $\beta$ Fermi surface splits into two sheets $\beta$$_1$ and $\beta$$_2$. It can also be seen directly from the measured band structure in Fig. 2: Cut 1 and Cut 2 in Fig. 2d, Cut 7 in Fig. 2f as well as Cut 6 and Cut 7 in Fig. 2h. The splitting is maximal along the (0,0)-($\pi$,0) and (0,0)-(0,$\pi$) directions but minimal along the (0,0)-($\pi$,$\pi$) direction, resulting in two split $\beta$ Fermi surface sheets as shown in Fig. 1n. The $\gamma$ band exhibits a similar splitting and the momentum dependence as the $\beta$ band. The $\gamma$ band splitting can be observed from the Fermi surface mapping in Fig. 1i where the $\gamma$ Fermi surface splits into two sheets $\gamma$$_1$ and $\gamma$$_2$. It can also be seen directly from the measured band structure in Fig. 2: Cut 6 and Cut 7 in Fig. 2h. The splitting is also maximal along (0,0)-(0,$\pi$) direction but minimal along the (0,0)-($\pi$,$\pi$) direction, resulting in two split $\gamma$ Fermi surface sheets as shown in Fig. 1n. Among all the measurements, we do not observe signature of band splitting for the $\alpha$ band. Overall, there are totally five hole pockets observed around $\Gamma$ point and tiny electron pockets observed around M point (Fig. 1n). The obtained hole concentration from the measured Fermi surface areas is 0.22$\pm$0.03 hole/Fe (the $\alpha$ pocket is considered twice in spite of its invisible band splitting). This concentration is consistent with the expected carrier concentration (0.25 hole/Fe) in stoichiometric K12442.

    The multiple Fermi surface sheets observed in K12442 provide a good opportunity to investigate the Fermi surface dependent superconducting gap and their overall momentum dependence. We start by examining the temperature dependence of the superconducting gap on all the Fermi surface sheets, as shown in Fig. 3. Fig. 3c-3f show the band structures measured at different temperatures along a momentum cut that covers $\alpha$, $\beta$$_1$ and $\beta$$_2$ bands; the location of the momentum cut is marked in Fig. 3a. These images are obtained by dividing the Fermi-Dirac distribution functions at their corresponding temperatures to facilitate investigation of the superconducting gap. In the normal state, all the observed bands smoothly cross the Fermi level (Fig. 3c). In the superconducting state, the spectral weight at the Fermi level gets suppressed due to the opening of the superconducting gap (Fig. 3d-3f). Fig. 3g-3i show the photoemmission spectra (Energy Distributed Curves, EDCs) measured at different temperatures on the Fermi momenta (marked by points in Fig. 3a) of $\alpha$, $\beta$$_1$ and $\beta$$_2$ bands, respectively. Sharp superconducting coherent peaks develop in the superconducting state for all the three bands. In order to quantitatively extract the superconducting gap size, the measured EDCs are symmetrized as shown in Fig. 3j-3l. The gap size is obtained by fitting these symmetrized EDCs with a phenomenological formula\cite{MRNorman1998DGHinks} and the obtained gap size at different temperatures for the three bands is plotted in Fig. 3t.

    Figure 3m-3n show the band structures measured at different temperatures along another momentum cut that covers $\gamma$$_1$ and $\gamma$$_2$ bands, as well as $\beta$$_1$ and $\beta$$_2$ bands. The location of the momentum cut is marked in Fig. 3b. Here the band splitting of the $\beta$ band and $\gamma$ band is particularly clear in the superconducting state (Fig. 3n). These two images are also obtained by dividing the Fermi-Dirac distribution functions at their corresponding temperatures. In the normal state, the observed two sets of bands smoothly cross the Fermi level (Fig. 3m). In the superconducting state, the spectral weight at the Fermi level gets suppressed due to the gap opening; it is more obvious for the $\beta$ band than that for the $\gamma$ band (Fig. 3n). Fig. 3o-3p show the EDCs measured at different temperatures on the Fermi momenta (marked by points in Fig. 3b) of $\gamma$$_1$ and $\gamma$$_2$ bands, respectively. The corresponding symmetrized EDCs are shown in Fig. 3q-3r and are fitted by the phenomenological formula\cite{MRNorman1998DGHinks}. The obtained superconducting gap size for the $\gamma$$_1$ and $\gamma$$_2$ bands is also included in Fig. 3t. Fig. 3s shows the temperature dependent symmetrized EDCs on the tiny electron pocket around M point. The extracted gap size by fitting the symmetrized EDCs is plotted in Fig. 3t. It can be found that superconducting gap opens on all the observed six bands and the gap size is different from each other (Fig. 3t). With increasing temperature, all the gaps close at around the superconducting transition temperature T$_C$.

    Now we come to the momentum dependence of the superconducting gap in this new superconductor K12442. Fig. 4 shows the measured superconducting gap for all the five hole-like pockets around the $\Gamma$ point. Taking advantage of the laser ARToF-ARPES system in simultaneous two-dimensional momentum coverage, the superconducting gap can be extracted from dense Fermi momentum points measured under the same condition. Fig. 4f-4j shows the symmetrized EDCs measured along the five hole-like Fermi surface sheets, $\alpha$, $\beta$$_1$, $\beta$$_2$, $\gamma$$_1$ and $\gamma$$_2$, respectively, in the supercondcuting state. Their corresponding Fermi surface and Fermi momentum location along the Fermi surface are given in Fig. 4a-4e. These symmetrized EDCs are fitted by the phenomenological formula\cite{MRNorman1998DGHinks} and the obtained superconducting gap size as a function of the Fermi surface angle for the $\alpha$, $\beta$$_1$ and $\beta$$_2$, $\gamma$$_1$ and $\gamma$$_2$ is shown in Fig. 4k-4m, respectively. Considering the fourfold symmetry of the crystal structure, we plot the superconducting gap distribution along the entire Fermi surface for the five hole-like pockets around $\Gamma$ and the electron pockets around M in Fig. 4n. The measured superconducting gap in K12442 exhibits a number of characteristics. First, the superconducting gap shows an obvious Fermi surface dependence; its size varies among different Fermi surface sheets. The maximal gap size ($\sim$8\,meV) appears on the $\beta$ Fermi surface sheets whereas the minimal gap size appears on the $\gamma$ sheets ($\sim$1\,meV). Second, the superconducting gap is nearly isotropic; it shows a slight anisotropy along the $\beta$$_2$ and $\gamma$ sheets. There is no gap node observed on all the Fermi surface sheets. Third, for the $\beta$$_1$ and $\beta$$_2$ Fermi surface sheets, their gap size is dramatically different although their location in the momentum space is very close.

    The multiple Fermi surface sheets observed in K12442 around $\Gamma$ point can be understood in terms of bilayer splitting similar to the one observed in the cuprate superconductor Bi2212. Bi2212 consists of two CuO$_2$ planes in one structural unit separated by calcium (Ca); the structural unit is separated by insulating block layers. It is highly two dimensional because it exhibits a very strong anisotropy between the out-of-plane and in-plane resistivities\cite{JVWaszczak1988SMartin}. Within the structural unit, the interaction between the two structurally equivalent CuO$_2$ planes gives rise to two Fermi surface sheets: bonding and antibonding ones\cite{FPaulsen1995OKAnderson,MLindroos1999ABansil,ZXShen2001PVBogdanov,DLFeng2001ZXShen,DSDessau2001YDChuang, TXiang2003YHSu}. In K12442, as shown in Fig. 1a, the bilayer FeAs block consists of two FeAs layers separated by K; this bilayer unit is separated by insulating Ca$_2$F$_2$ block\cite{GHCao2016ZCWang}. It also shows strong anisotropy between the out-of-plane and in-plane resistivities\cite{GHCao2019ZCWang}. Within the bilayer FeAs block, the interaction between the two FeAs layers will cause the band splitting into two bands for each of the initial three bands in the single FeAs layer. The band splittings we observed in K12442 for the $\beta$ and $\gamma$ bands can be understood in this bilayer splitting picture, similar to that in Bi2212. It is not observed for the $\alpha$ band because the bilayer splitting is Fermi surface dependent and the splitting is too small to be detected for the $\alpha$ band.

    We carried out band structure calculations to further understand the interaction between the two adjacent FeAs layers and the microscopic origin of the bilayer splitting in K12442. In the calculations, we adopted five-band tight-binding model for each FeAs layer and included the interlayer hopping for $d_{xz/yz}$ and $d_{xy}$ orbitals up to the third nearest neighbor (see the supplementary materials). Fig. 2i shows the calculated Fermi surface; the corresponding calculated band structures along several momentum cuts are shown in Fig. 2j. Compared with experimental data in Fig. 2c-2h, the band splitting, its Fermi surface dependence and the momentum dependence are well reproduced by the calculations. We found that the interlayer interorbital ($d_{xz}$ and $d_{yz}$) hopping along the body diagonal (third nearest neighbour interaction t$_2$) in Fe lattice, as shown in Fig. 2k, plays the dominant role in understanding the observed momentum-dependent band splitting of the $\beta$ band. The $\alpha$ band shares similar orbital components and the interlayer interorbital interaction (Fig. 2k), however, its band splitting is invisible because it is too small due to the small momentum radius of its Fermi surface (explained in supplementary materials). In order to understand the band splitting of the $\gamma$ band, both the nearest (t$_0$) and the third nearest (t$_2$) neighbour interlayer interorbital interactions have to be considered, as shown in Fig. 2l.

    K12442 has the largest number of observed Fermi surface sheets among all the iron based superconductors discovered so far. With such multiple Fermi surface sheets, together with the detailed momentum dependent superconducting gap measured on each sheet, K12442 provides an ideal system to examine on possible gap functions and explore on possible pairing mechanisms. Generally, there are two approaches to understand superconductivity of the iron based superconductors\cite{IIMazin2011PJHirschfeld,DHLee2011FWang}. In the Fermi surface nesting scenario of the itinerant picture\cite{HAoki2008KKuroki,MHDu2008IIMazin,IIMazin2011PJHirschfeld,DHLee2011FWang}, the electron scattering between the hole pockets around $\Gamma$ and the electron pockets around M is proposed to be responsible for electron pairing. Such a mechanism was considered in (Ba$_{0.6}$K$_{0.4}$)Fe$_2$As$_2$\cite{NLWang2008HDing} and CaKFe$_4$As$_4$\cite{AKaminsk2016DXMou} superconductors. The possibility of this nesting-driven pairing mechanism can be ruled out in K12442 because of the apparent Fermi surface size mismatch between the hole pockets around $\Gamma$ and the electron pockets around M, as seen in Fig. 1n.

    For the strong coupling approach\cite{JPHu2008KSeo,IIMazin2011PJHirschfeld,DHLee2011FWang}, the pairing of electrons occurs because of a short-range interaction. Referring to (Ba$_{0.6}$K$_{0.4}$)Fe$_2$As$_2$\cite{HDing2011YMXu}, we propose a generalized $s$-wave gap function for K12442: $\Delta_s=|\frac{1}{2}\Delta_0(cosk_x+cosk_y)\pm\frac{1}{2}\Delta_zcos\frac{k_x}{2}cos\frac{k_y}{2}|$ in 2-Fe unit cell, where the intralayer pairing $\Delta_0$ originates from the intralayer next-nearest neighbor exchange coupling $J_2$ and the interlayer pairing $\Delta_z$ comes from interlayer exchange coupling $J_z$ (see inset in Fig. 4p and the supplementary materials). First, $\Delta_z$ has to be considered in order to understand the dramatic difference of the superconducting gap for the two $\beta$ split bands. Second, if we consider the interlayer pairing but assume band-independent $\Delta_0$ and $\Delta_z$, clear deviation still exists, as shown in Fig. 4o, particularly for the $\beta_2$ and $\gamma_1$ bands. Only when we consider the interlayer pairing and band-dependent $\Delta_0$ and $\Delta_z$ can we fit the observed gap well, as shown in Fig. 4p. The $\Delta_0$ and $\Delta_z$ for different Fermi surface sheets are thus obtained, as also included in Fig. 4p. We find that the intralayer pairing shows a clear Fermi surface dependence; it is strongest for the $\beta$ band ($\Delta_0$=6.8\,meV) while weakest for the $\gamma$ band ($\Delta_0$=3.3\,meV). Furthermore, the interlayer pairing also exhibits a Fermi surface dependence which is also strongest for the $\beta$ band ($\Delta_z$=4\,meV). The interlayer pairing for the $\beta$ band is stronger than that observed in (Ba$_{0.6}$K$_{0.4}$)Fe$_2$As$_2$ ($\Delta_z$=2.07\,meV)\cite{HDing2011YMXu} and is the strongest discovered so far in iron based superconductors.

    In summary, we carried on high resolution laser ARPES measurements on a newly discovered iron based superconductor, KCa$_2$Fe$_4$As$_4$F$_2$. Totally five hole-like Fermi surface sheets around $\Gamma$ point and tiny electron-like pockets around M point are observed. The bilayer splitting effect is clearly identified which represents the first case discovered in iron based superconductors. It mainly originates from the interlayer interorbital interactions. Fermi surface-dependent and nodeless superconducting gap is observed. In particular, the superconducting gap on the two split $\beta$ bands is dramatically different and the maximal gap size appears on one of the bands. The observed Fermi surface topology clearly rules out the Fermi surface nesting picture in understanding the superconductivity in KCa$_2$Fe$_4$As$_4$F$_2$.  Instead, the measured superconducting gap is consistent with the gap functions considering short-range and band dependent pairing; the particularly strong interlayer pairing is observed for the $\beta$ band. Our results provide key insights in the role of the interlayer interaction and interlayer pairing played in dictating the superconductivity in iron based superconductors.

\vspace{3mm}

\vspace{3mm}

\vspace{3mm}

\vspace{3mm}

$^{*}$Corresponding author: LZhao@iphy.ac.cn, hqluo@aphy.iphy.ac.cn, jphu@iphy.ac.cn, XJZhou@iphy.ac.cn.

\bibliographystyle{unsrt}
\bibliography{IronSCRef}

\begin{thebibliography}{10}

\bibitem{MGreven2004HEisaki}
{H. Eisaki, N. Kaneko, D. L. Feng, A. Damascelli, P. K. Mang, K. M. Shen, Z. X.
  Shen and M. Greven}.
\newblock {Effect of Chemical Inhomogeneity in Bismuth-Based Copper Oxide
  Superconductors}.
\newblock {\em Physical Review B}, 69(6):064512, 2004.

\bibitem{ZXShen2001PVBogdanov}
{P. V. Bogdanov, A. Lanzara, X. J. Zhou, S. A. Kellar, D. L. Feng, E. D. Lu, H.
  Eisaki, J. I. Shimoyama, K. Kishio, Z. Hussain and Z. X. Shen}.
\newblock {Photoemission Study of Pb Doped Bi$_2$Sr$_2$CaCu$_2$O$_8$: A Fermi
  Surface Picture}.
\newblock {\em Physical Review B}, 64(18):180505(R), 2001.

\bibitem{DLFeng2001ZXShen}
{D. L. Feng, N. P. Armitage, D. H. Lu, A. Damascelli, J. P. Hu, P. Bogdanov, A.
  Lanzara, F. Ronning, K. M. Shen, H. Eisaki, C. Kim, Z. X. Shen, J. Shimoyama
  and K. Kishio}.
\newblock {Bilayer Splitting in the Electronic Structure of Heavily Overdoped
  Bi$_2$Sr$_2$CaCu$_2$O$_{8+\delta}$}.
\newblock {\em Physical Review Letters}, 86(24):5550--3, 2001.

\bibitem{DSDessau2001YDChuang}
{Y. D. Chuang, A. D. Gromko, A. Fedorov, Y. Aiura, K. Oka, Y. Ando, H. Eisaki,
  S. I. Uchida and D. S. Dessau}.
\newblock {Doubling of the Bands in Overdoped
  Bi$_2$Sr$_2$CaCu$_2$O$_{8+\delta}$: Evidence for c-axis Bilayer Coupling}.
\newblock {\em Physical Review Letters}, 87(11):117002, 2001.

\bibitem{XJZhou2019PAi}
{P. Ai, Q. Gao, J. Liu, Y. X. Zhang, C. Li, J. W. Huang, C. Y.
  Song, H. T. Yan, L. Zhao, G. D. Liu, G. D. Gu, F. F. Zhang, F.
  Yang, Q. J. Peng, Z. Y. Xu and X. J. Zhou}.
\newblock {Distinct Superconducting Gap on Two Bilayer-Split Fermi Surface
  Sheets in Bi$_2$Sr$_2$CaCu$_2$O$_{8+\delta}$ Superconductor}.
\newblock {\em Chinese Physics Letters}, 36(6):067402, 2019.

\bibitem{GHCao2016ZCWang}
{Z. C. Wang, C. Y. He, S. Q. Wu, Z. T. Tang, Y. Liu, A. Ablimit, C. M. Feng and
  G. H. Cao}.
\newblock {Superconductivity in KCa$_2$Fe$_4$As$_4$F$_2$ with Separate Double
  Fe$_2$As$_2$ Layers}.
\newblock {\em Journal of the American Chemical Society}, 138(25):7856--9,
  2016.

\bibitem{GHCao2017ZCWang}
{Z. C. Wang, C. Y. He, Z. T. Tang, S. Q. Wu and G. H. Cao}.
\newblock {Crystal Structure and Superconductivity at about 30\,K in
  ACa$_2$Fe$_4$As$_4$F$_2$ (A = Rb, Cs)}.
\newblock {\em Science China Materials}, 60(1):83--89, 2016.

\bibitem{XBShi2016GTWang}
{G. T. Wang, Z. W. Wang and X. B. Shi}.
\newblock {Self-Hole-Doping Induced Superconductivity in
  KCa$_2$Fe$_4$As$_4$F$_2$}.
\newblock {\em Europhysics Letters}, 116(3):37003, 2016.

\bibitem{ADHillier2017JIshida}
{J. Ishida, S. Iimura and H. Hosono}.
\newblock {Effects of Disorder on the Intrinsically Hole-Doped Iron-Based
  Superconductor KCa$_2$Fe$_4$As$_4$F$_2$ by Cobalt Substitution}.
\newblock {\em Physical Review B}, 96(17):174522, 2017.

\bibitem{SJBlundell2018FKKKirschner}
{F. K. K. Kirschner, D. T. Adroja, Z. C. Wang, F. Lang,
  M. Smidman, P. J. Baker, G. H. Cao and S. J. Blundell}.
\newblock {Two-Gap Superconductivity with Line Nodes in
  CsCa$_2$Fe$_4$As$_4$F$_2$}.
\newblock {\em Physical Review B}, 97(6):060506(R), 2018.

\bibitem{SJBlundell2018MSmidman}
{M. Midman, F. K. K. Kirschner, D. T. Adroja, A. D. Hillier, F. Lang, Z. C. Wang,
  G. H. Cao and S. J. Blundell}.
\newblock {Nodal Multigap Superconductivity in KCa$_2$Fe$_4$As$_4$F$_2$}.
\newblock {\em Physical Review B}, 97(6):060509(R), 2018.

\bibitem{CBernhard2019BXu}
{B. Xu, Z. C. Wang, E. Sheveleva, F. Lyzwa, P. Marsik, G. H. Cao and C.
  Bernhard}.
\newblock {Band-Selective Clean-Limit and Dirty-Limit Superconductivity with
  Nodeless Gaps in the Bilayer Iron-Based Superconductor
  CsCa$_2$Fe$_4$As$_4$F$_2$}.
\newblock {\em Physical Review B}, 99(12):125119, 2019.

\bibitem{SYLi2019YYHuang}
{Y. Y. Huang, Z. C. Wang, Y. J. Yu, J. M. Ni, Q. Li, E. J. Cheng, G. H. Cao and
  S. Y. Li}.
\newblock {Multigap Nodeless Superconductivity in CsCa$_2$Fe$_4$As$_4$F$_2$
  Probed by Heat Transport}.
\newblock {\em Physical Review B}, 99(2):020502(R), 2019.

\bibitem{GMu2019TWang}
{T. Wang, J. N. Chu, J. X. Feng, L. L. Wang, X. G. Xu, W. Li, H. H.
  Wen, X. S. Liu and G. Mu}.
\newblock {Low Temperature Specific Heat of the 12442-Type
  KCa$_2$Fe$_4$As$_4$F$_2$ Single Crystals}.
\newblock {\em cond-mat arXiv}  :1903.09447, 2019.

\bibitem{GHCao2019ZCWang}
{Z. C. Wang, Y. Liu, S. Q. Wu, Y. T. Shao, Z. Ren and G. H. Cao}.
\newblock {Giant Anisotropy in Superconducting Single Crystals of
  CsCa$_2$Fe$_4$As$_4$F$_2$}.
\newblock {\em Physical Review B}, 99(14):144501, 2019.

\bibitem{HHosono2008YKamihara}
Y. Kamihara, T. Watanabe, M. Hirano and H. Hosono.
\newblock {Iron-Based Layered Superconductor La[O$_{1-x}$F$_x$]FeAs
  (x=0.05-0.12) with T$_c$=26 K}.
\newblock {\em Journal of the American Chemical Society}, 130:3296--3297, 2008.

\bibitem{MWu2008FHsu}
{F. Hsu, J. Luo, K. Yeh, T. Chen, T. Huang, P. M. Wu, Y. Lee, Y. Huang, Y. Chu,
  D. Yan and M. Wu}.
\newblock {Superconductivity in the PbO-Type Structure $\alpha$-FeSe}.
\newblock {\em Proceedings of the National Academy of Sciences of the United
  State of America}, 105:14262--14264, 2008.

\bibitem{CQJin2008XCWang}
{X. C. Wang, Q. Q. Liu, Y. X. Lv, W. B. Gao, L. X. Yang, R. C. Yu, F. Y. Li and
  C. Q. Jin}.
\newblock {The Superconductivity at 18 K in LiFeAs System}.
\newblock {\em Solid State Communications}, 148(11-12):538--540, 2008.

\bibitem{DJohrendt2008MRotter}
M. Rotter, M. Tegel, D. Johrendt, I. Schellenberg, W.
  Hermes and R. Pottgen.
\newblock Spin-Density-Wave Anomaly at 140 K in the Ternary Iron Arsenide
  BaFe$_2$As$_2$.
\newblock {\em Physical Review B},
  78(2):020503--4, 2008.

\bibitem{RBernard2013MClaude}
B. Raveau, C. Michel, M. Hervieu and D. Groult.
\newblock {\em Crystal Chemistry of High-Tc Superconducting Copper Oxides},
  volume~15.
\newblock Springer Science \& Business Media, 2013.

\bibitem{XJZhou2018}
{X. J. Zhou, S. L. He, G. D. Liu, L. Zhao, L. Yu and W. T. Zhang}.
\newblock {New Developments in Laser-Based Photoemission Spectroscopy and Its
  Scientific Applications: A Key Issues Review}.
\newblock {\em Reports on Progress in Physics}, 81(6):062101, 2018.

\bibitem{XJZhou2017YZhangNC}
  Y. Zhang, C. L. Wang, L. Yu, G. D. Liu, A. J. Liang, J. W. Huang, S. M.
  Nie, X. Sun, Y. X. Zhang, B. Shen, J. Liu, H. M. Weng, L. X.
  Zhao, G. F. Chen, X. W. Jia, C. Hu, Y. Ding, W. J. Zhao, Q. Gao,
  C. Li, S. L. He, L. Zhao, F. F. Zhang, S. J. Zhang, F. Yang,
  Z. M. Wang, Q. J. Peng, X. Dai, Z. Fang, Z. Y. Xu, C. T. Chen and
  X. J. Zhou.
\newblock {Electronic Evidence of Temperature-Induced Lifshitz Transition and
  Topological Nature in ZrTe$_5$}.
\newblock {\em Nature Communications}, 8:15512, 2017.

\bibitem{Damascelli2003}
A. Damascelli, Z. Hussain and Z. X. Shen.
\newblock {Angle-Resolved Photoemission Studies of the Cuprate
  Superconductors}.
\newblock {\em Reviews of Modern Physics}, 75(2):473--541, 2003.

\bibitem{MRNorman1998DGHinks}
{M. R. Norman, H. Ding, M. Randeria, J. C. Campuzano, T. Yokoya, T. Takeuchik,
  T. Takahashi, T. Mochiku, K. Kadowaki, P. Guptasarma and D. G. Hinks}.
\newblock {Destruction of the Fermi Surface in Underdoped High-Tc
  Superconductors}.
\newblock {\em Nature}, 392:157--160, 1998.

\bibitem{JVWaszczak1988SMartin}
{S. Martin, A. T. Fiory, R. M. Fleming, L. F. Schneemeyer and J. V. Waszczak}.
\newblock {Temperature Dependence of the Resistivity Tensor in Superconducting
  Bi$_2$Sr$_{2.2}$Ca$_{0.8}$Cu$_2$O$_8$ Crystals}.
\newblock {\em Physical Review Letters}, 60(21):2194--2197, 1988.

\bibitem{FPaulsen1995OKAnderson}
{O. K. Andersen, A. I. Liechtenstein, O. Jepsen and F. Paulsen}.
\newblock {LDA Energy Bands, Low-Energy Hamiltonians,
  t$^\prime$,t$^\prime$$^\prime$,t$_\bot$(k) and J$_\bot$}.
\newblock {\em Journal of Physics and Chemistry of Solids}, 56:1573--1594, 1995.

\bibitem{MLindroos1999ABansil}
A. Bansil and M. Lindroos.
\newblock Importance of Matrix Elements in the ARPES Spectra of BISCO.
\newblock {\em Physical Review Letters}, 83:5154, 1999.

\bibitem{TXiang2003YHSu}
{Y. H. Su, J. Chang, H. T. Lu, H. G. Luo and T. Xiang}.
\newblock {Effect of Bilayer Coupling on Tunneling Conductance of Double-Layer
  High-T$_c$ Cuprates}.
\newblock {\em Physical Review B}, 68(21):212501, 2003.

\bibitem{IIMazin2011PJHirschfeld}
{P. J. Hirschfeld, M. M. Korshunov and I. I. Mazin}.
\newblock {Gap Symmetry and Structure of Fe-Based Superconductors}.
\newblock {\em Reports on Progress in Physics}, 74(12):124508, 2011.

\bibitem{DHLee2011FWang}
{F. Wang and D. H. Lee}.
\newblock {The Electron-Pairing Mechanism of Iron-Based Superconductors}.
\newblock {\em Science}, 332:200--204, 2011.

\bibitem{HAoki2008KKuroki}
{K. Kuroki, S. Onari, R. Arita, H. Usui,
  Y. Tanaka, H. Kontani and H. Aoki}.
\newblock {Unconventional Pairing Originating from the Disconnected Fermi
  Surfaces of Superconducting LaFeAsO$_{1-x}$F$_x$}.
\newblock {\em Physical Review Letters}, 101(8):087004, 2008.

\bibitem{MHDu2008IIMazin}
I. I. Mazin, D. J. Singh, M. D. Johannes and M. H. Du.
\newblock {Unconventional Superconductivity with a Sign Reversal in the Order
  Parameter of LaFeAsO$_{1-x}$F$_x$}.
\newblock {\em Physical Review Letters}, 101(5):057003--4, 2008.

\bibitem{NLWang2008HDing}
H. Ding, P. Richard, K. Nakayama, K. Sugawara, T. Arakane, Y. Sekiba,
  A. Takayama, S. Souma, T. Sato, T. Takahashi, Z. Wang, X. Dai, Z. Fang, G. F.
  Chen, J. L. Luo and N. L. Wang.
\newblock {Observation of Fermi-Surface-Dependent Nodeless Superconducting Gaps
  in Ba$_{0.6}$K$_{0.4}$Fe$_2$As$_2$}.
\newblock {\em Europhysics Letters}, 83(4):47001, 2008.

\bibitem{AKaminsk2016DXMou}
{D. X. Mou, T. Kong, W. R. Meier, F. Lochner, L. L. Wang, Q. S. Lin, Y. Wu, S. L.
  Bud'ko, I. Eremin, D. D. Johnson, P. C. Canfield and A. Kaminski}.
\newblock {Enhancement of the Superconducting Gap by Nesting in
  CaKFe$_4$As$_4$: A New High Temperature Superconductor}.
\newblock {\em Physical Review Letters}, 117(27):277001, 2016.

\bibitem{JPHu2008KSeo}
{K. Seo, B. A. Bernevig and J. P. Hu}.
\newblock {Pairing Symmetry in a Two-Orbital Exchange Coupling Model of
  Oxypnictides}.
\newblock {\em Physical Review Letters}, 101(20):206404, 2008.

\bibitem{HDing2011YMXu}
Y. M. Xu, Y. B. Huang, X. Y. Cui, E. Razzoli, M. Radovic, M. Shi, G. F. Chen,
  P. Zheng, N. L. Wang, C. L. Zhang, P. C. Dai, J. P. Hu, Z. Wang and H. Ding.
\newblock {Observation of a Ubiquitous Three-Dimensional Superconducting Gap
  Function in Optimally Doped Ba$_{0.6}$K$_{0.4}$Fe$_2$As$_2$}.
\newblock {\em Nature Physics}, 7(3):198--202, 2011.

\end{thebibliography}



\vspace{3mm}

\noindent {\bf Acknowledgement}\\
 This work is supported by the National Natural Science Foundation of China (Grant No. 11888101, 11534007, 11822411, 11374011, 11922414 and 11874405), the National Key Research and Development Program of China (Grant No. 2016YFA0300300, 2017YFA0302900, 2018YFA0704200 and 2019YFA0308000), the Strategic Priority Research Program (B) of the Chinese Academy of Sciences (Grant No. XDB25000000), the Research Program of Beijing Academy of Quantum Information Sciences (Grant No. Y18G06) and the Youth Innovation Promotion Association of CAS (Grant No. 2017013, 2016004 and 2019007).

\vspace{3mm}

\noindent {\bf Author Contributions}\\
 X.J.Z., L.Z. and D.S.W. proposed and designed the research. W.S.H., H.Q.L. and S.L.L. contributed in sample growth. D.S.W., J.W.H., Q.G., C.L., C.Y.S., H.L.L., C.H.Y., Y.X., X.Y.L., Y.Q.C., J.J.J., Q.Y.W., Y.H., G.D.L., S.J.Z., F.F.Z., F.Y., Z.M.W., Q,J,P., Z.Y.X., L.Z. and X.J.Z. contributed to the development and maintenance of the ARPES systems and related software development. Q.T.S., X.G.Q., W.S.H, H.Q.L. and D.S.W. contributed to the magnetic and resistivity measurements. D.S.W. carried out the ARPES experiment with L.Z.. D.S.W., L.Z. and X.J.Z. analyzed the data. C.X.D., X.X.W. and J.P.H. contributed to the band structure calculations and theoretical analysis. D.S.W., L.Z. and X.J.Z. wrote the paper with C.X.D., X.X.W. and J.P.H.. All authors participated in discussion and comment on the paper.\\

\vspace{3mm}

\noindent{\bf Additional information}\\
Supplementary information is available in the online version of the paper.
Correspondence and requests for materials should be addressed to L. Zhao, H. Q. Luo, J. P. Hu and X. J. Zhou.

\newpage

\begin{figure*}[tbp]
\begin{center}
\includegraphics[width=1\columnwidth,angle=0]{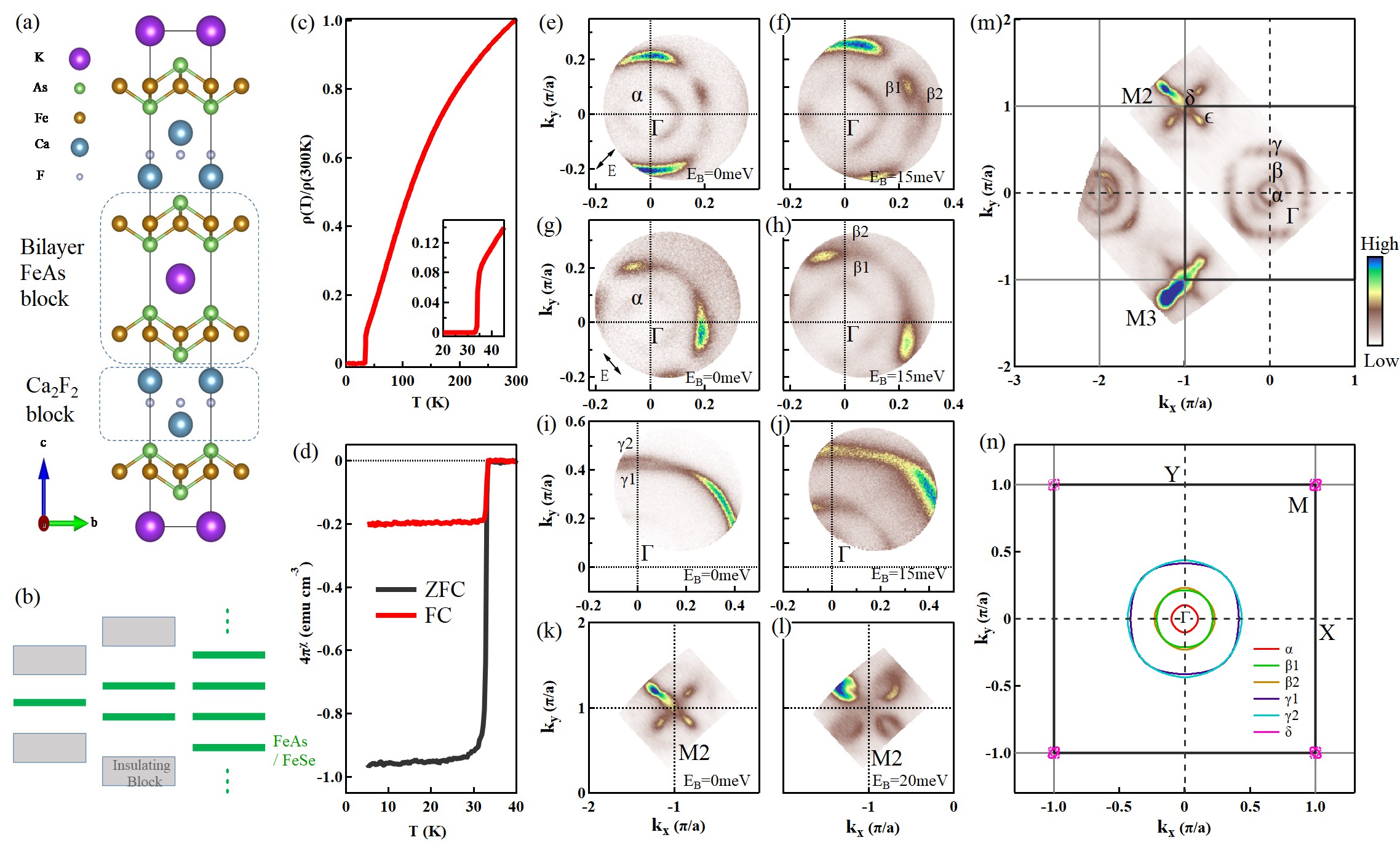}
\end{center}

\caption{{\bf Crystal structure, superconducting properties and measured Fermi surface of K12442 superconductor with a T$_c$ of 33.5\,K.} (a) Crystal structure consisting of bilayer FeAs block separated by insulating Ca$_2$F$_2$ block. (b) Schematic crystal structures of three groups of iron based superconductors. Left panel: single FeAs/FeSe layer separated by insulating blocks; middle panel: double FeAs/FeSe layers separated by insulating blocks; right panel: infinite stacking of FeAs/FeSe layers. (c) Resistivity measurement within $\it ab$-plane. The inset shows the data near the superconducting transition. (d) Magnetic susceptibility measurement with zero-field-cooling (ZFC) and field-cooling (FC) methods under a magnetic field of 0.9\,Oe. (e-l) Fermi surface mappings and constant energy contours covering different momentum space. The measurements around $\Gamma$ point were carried out by using laser at 12-13\,K for (e)-(j) and the measurements around M point were performed by using helium lamp at 38\,K for (k)-(l). (e,g,i,k) represents Fermi surface mappings and (f,h,j,l) shows their corresponding constant energy contours at a binding energy of 15\,meV (f,h,j) or 20\,meV (l). They are obtained by integrating the spectral weight within a narrow energy window with respect to the Fermi level or the corresponding binding energies. The black double arrows in (e) and (g) represent E vector of the incident laser light. The observed Fermi surface sheets are marked by $\alpha$, $\beta_1$, $\beta_2$, $\gamma_1$ and $\gamma_2$. The splitting of the $\beta$ Fermi surface can be seen in (f), (h) and (j), and the splitting of the $\gamma$ Fermi surface can be seen in (i). (m) Overall Fermi surface mapping consisting of three hole-like pockets ($\alpha$, $\beta$ and $\gamma$) around $\Gamma$ point and small electron-like pockets ($\delta$) around M point measured at 38\,K in the normal state with the helium lamp. (n) Summary of all the observed Fermi surface sheets obtained from the data in (e-m).
}
\end{figure*}
\pagestyle{empty}

\begin{figure*}[tbp]
\begin{center}
\includegraphics[width=1.0\columnwidth,angle=0]{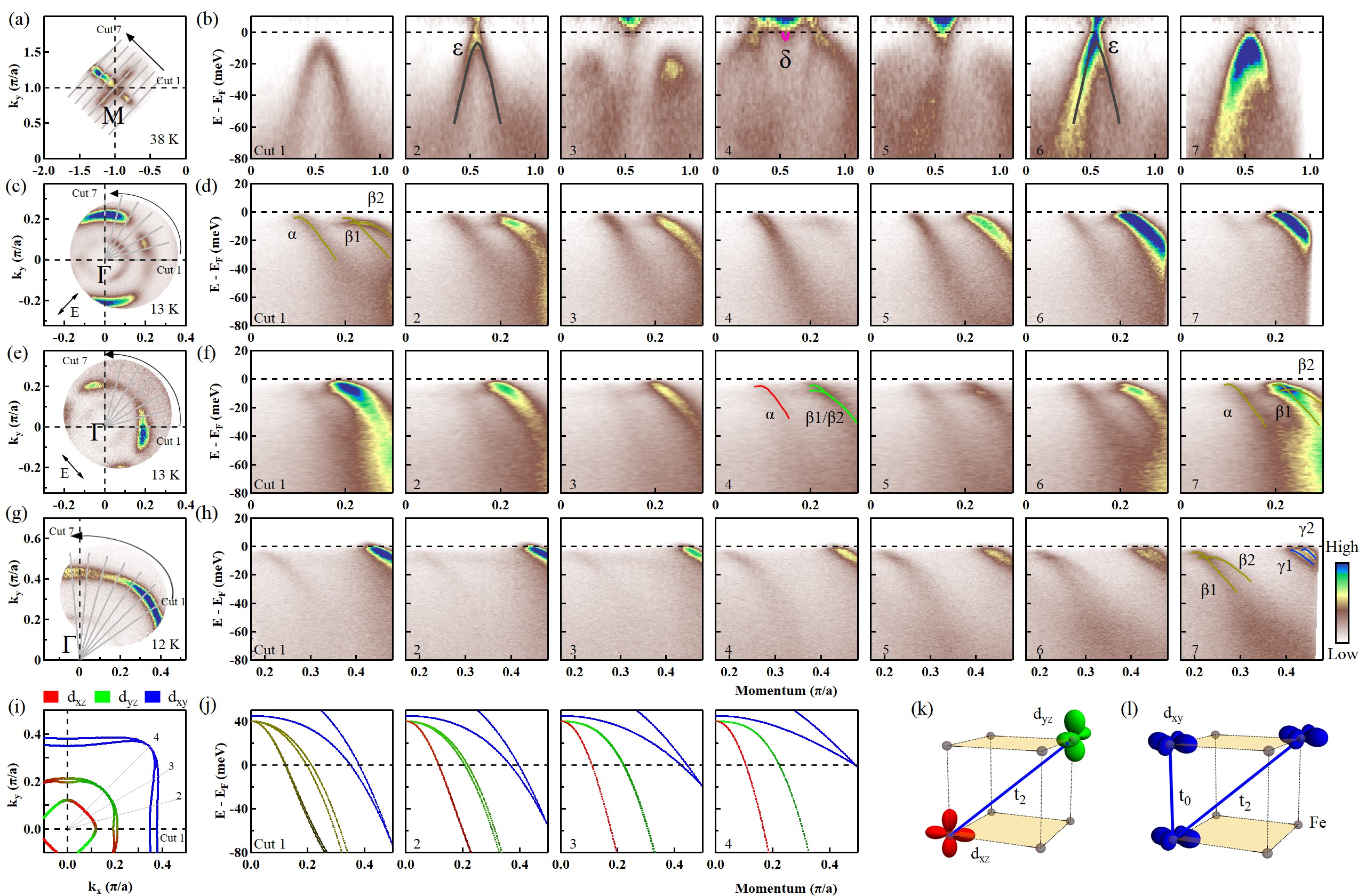}
\end{center}

\caption{{\bf Momentum dependent band structures of K12442 superconductor and the comparison with band calculations.} (a) Fermi surface mapping around M point measured by using helium lamp at a temperature of 38\,K. The location of seven momentum cuts are marked by grey lines and their corresponding band structures are shown in (b). In order to see the features above the Fermi level, these spectral images are obtained by dividing the original data with the Fermi-Dirac distribution function. The $\epsilon$ band for the cuts 2 and 6 crossing the strong spots and the $\delta$ band for the cut 4 crossing the M point are marked. Similar to (a), (c,e,g) show Fermi surface mappings around $\Gamma$ point measured by using laser at 12-13\,K in the superconducting state and the location of the momentum cuts. The corresponding band structures are shown in (d,f,h). The observed bands, $\alpha$, $\beta$$_1$ and $\beta$$_2$, $\gamma$$_1$ and $\gamma$$_2$, are marked in some of the images. (i) Calculated Fermi surface by considering the interlayer interaction between the two FeAs layers inside the bilayer FeAs block. The calculated momentum dependent band structures are shown in (j); the location of the momentum cuts is marked as grey lines in (i). The d$_{xz}$, d$_{yz}$, and d$_{xy}$ orbital components are represented by red, green and blue colors, respectively. (k) Schematic of the main interlayer interorbital interaction involving the second next nearest neighbour hopping (t$_2$) and d$_{xz/yz}$ orbitals that causes the splitting of the $\beta$ band. (l) Schematic of the main interlayer interorbital interaction involving the nearest (t$_0$) and the second nearest neighbour (t$_2$) hoppings and d$_{xy}$ orbital that causes the splitting of the $\gamma$ band.
}
\end{figure*}

\begin{figure*}[tbp]
\begin{center}
\includegraphics[width=1.0\columnwidth,angle=0]{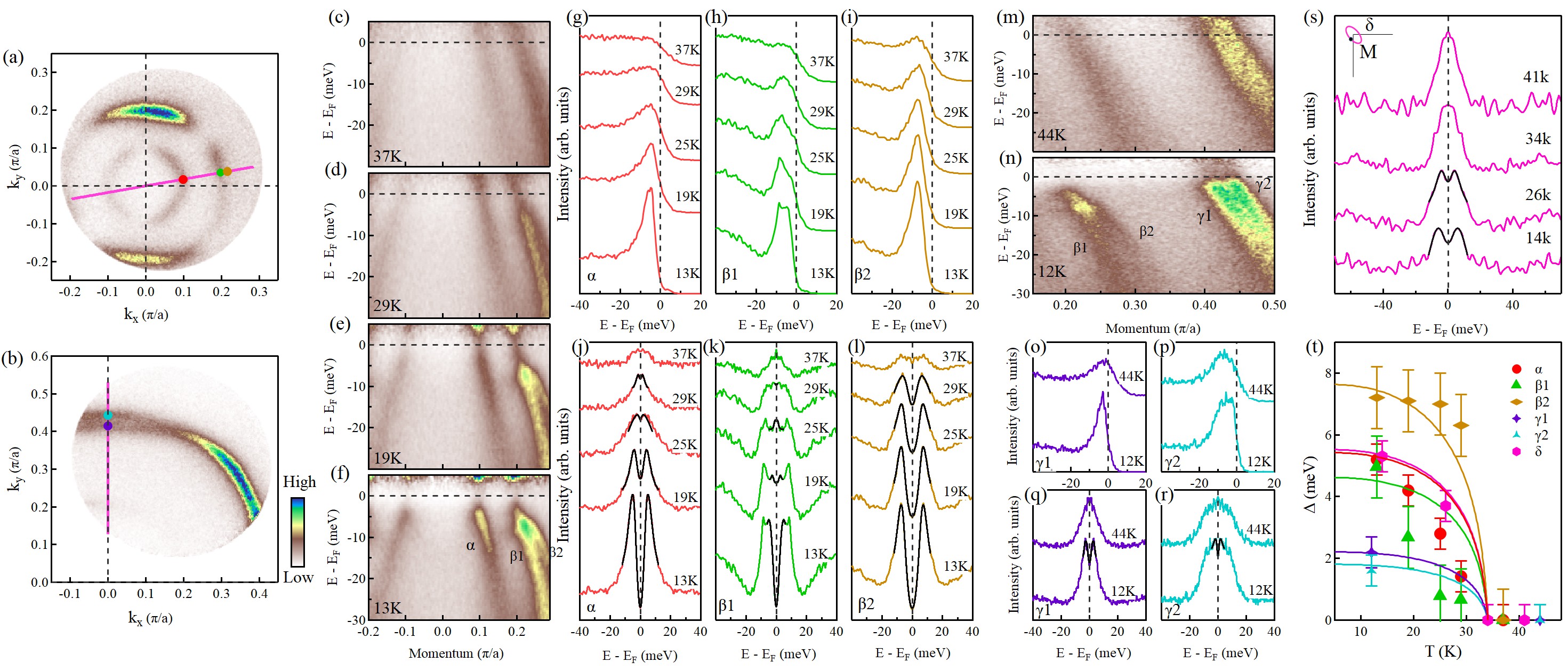}
\end{center}

\caption{{\bf Temperature dependence of superconducting gap in K12442 superconductor.} (a) Fermi surface mapping measured at 13\,K by using laser. It covers $\alpha$, $\beta$$_1$ and $\beta$$_2$ Fermi pockets. The pink line marks the position of a momentum cut and the red, green and orange points represent the Fermi momentum position of the $\alpha$, $\beta$$_1$ and $\beta$$_2$ Fermi surface sheets. (b) Same as (a) but covers $\beta$$_1$, $\beta$$_2$, $\gamma$$_1$ and $\gamma$$_2$ Fermi pockets. The pink line marks the position of another momentum cut and the purple and blue points represent the Fermi momentum position of the $\gamma$$_1$ and $\gamma$$_2$ Fermi surface sheets. (c-f) Band structures measured at four different temperatures along the momentum cut shown in (a). The spectral images are divided by the Fermi-Dirac distribution function at their corresponding temperatures in order to highlight the superconducting gap. (g-i) Photoemmission spectra (EDCs) measured at different temperatures for the three Fermi momenta, as indicated in (a), on $\alpha$, $\beta$$_1$ and $\beta$$_2$ pockets, respectively. The corresponding symmetrized EDCs are shown in (j-l). (m-n) Band structures measured at two different temperatures along the momentum cut shown in (b). The spectral images are also divided by the Fermi-Dirac distribution function at their corresponding temperatures. (o-p) EDCs measured at different temperatures for the two Fermi momenta, as indicated in (b), on $\gamma$$_1$ and $\gamma$$_2$ pockets, respectively. The corresponding symmetrized EDCs are shown in (q-r). (s) Symmetrized EDCs at the Fermi momentum of the tiny electron pocket around M point measured at different temperatures by using helium lamp. All the symmetrized EDCs in (j,k,l,q,r,s) are fitted by the phenomenological formula\cite{MRNorman1998DGHinks}; the obtained superconducting gap size as a function of temperature for all the six Fermi surface sheets is plotted in (t). The solid lines represent curves corresponding to the BCS gap function. Superconducting gap opens on all the Fermi surface sheets in the superconducting state and closes at the superconducting transition temperature.
   }
\end{figure*}

\begin{figure*}[tbp]
\begin{center}
\includegraphics[width=1.0\columnwidth,angle=0]{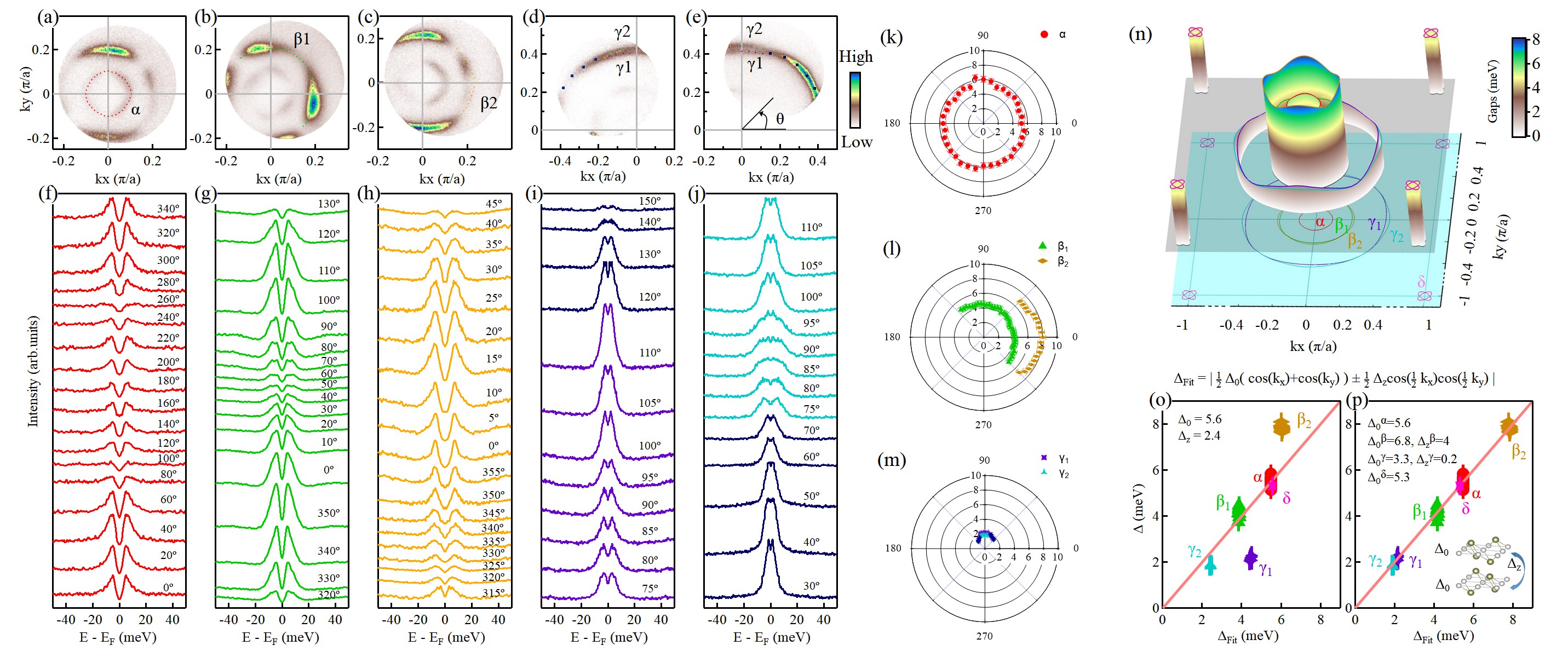}
\end{center}

\caption{{\bf Momentum-dependent superconducting gap of K12442 superconductor.} (a-e) Fermi surface mappings measured at low temperature of 12-13\,K. (f-j) Symmetrized EDCs along the $\alpha$, $\beta$$_1$, $\beta$$_2$, $\gamma$$_1$ and $ \gamma$$_2$ Fermi surface sheets, respectively, as a function of the Fermi surface angle $\theta$. The Fermi surface angle $\theta$ is defined with respect to the $\Gamma$-X direction, as shown in (e). The location of the Fermi momenta is also marked in the corresponding Fermi surface mappings. The symmetrized EDCs of $\gamma$$_1$ and $\gamma$$_2$ in (i) and (j) are extracted from both (d) and (e). All the symmetrized EDCs in (f-j) are fitted by the phenomenological formula\cite{MRNorman1998DGHinks}; the obtained superconducting gap size as a function of the Fermi surface angle for the $\alpha$, $\beta$$_1$ and $\beta$$_2$, $\gamma$$_1$ and $\gamma$$_2$ Fermi surface sheets is plotted as polar graphs in (k), (l) and (m), respectively. (n) Three-dimensional plot of the superconducting gap on the five observed hole pockets around $\Gamma$ point and the tiny electron pockets around M point. The corresponding Fermi surface is shown at the bottom. (o) Superconducting gap size on various Fermi surface sheets as a function of $\Delta_s=|\frac{1}{2}\Delta_0(cosk_x+cosk_y)\pm\frac{1}{2}\Delta_zcos\frac{k_x}{2}cos\frac{k_y}{2}|$ by considering Fermi surface-independent intralayer pairing ($\Delta_0$) and interlayer pairing ($\Delta_z$). (p) Same as (o) but considering Fermi surface-dependent intralayer pairing ($\Delta_0$) and interlayer pairing ($\Delta_z$).
   }

\end{figure*}

\end{document}


\title{Supplementary Materials for\\
Spectroscopic Evidence of Bilayer Splitting and Interlayer Pairing in an Iron Based Superconductor\\}

\date{January 13, 2020}

\maketitle

%
\section{Tight-Binding Band Calculations}
To simulate the band structure of K12442 system, we adopted the bilayer five-orbital tight-binding model with the interlayer hopping term
\begin{equation}
\begin{split}
H_{interlayer}=&2t_2^{xz,yz}\sin{kx}\sin{ky}(c^{+}_{1,xz}c_{2,yz}+c^{+}_{1,yz}c_{2,xz}+c^{+}_{2,xz}c_{1,yz}+c^{+}_{2,yz}c_{1,xz})\\&+(t_0^{xy}+2t_2^{xy}\cos{kx}\cos{ky})(c^{+}_{1,xy}c_{2,xy}+c^{+}_{2,xy}c_{1,xy})
\end{split}
\end{equation}
where $c^{+}_{1,xz}$ is the creation operator of the orbit $d_{xz}$ in layer 1, the rest operators are in the same way; $t_2^{xz,yz}=20\,meV$; $t_0^{xy}=-14\,meV$; $t_2^{xy}=15\,meV$.

\setlength{\parindent}{2em} From equation (1), it is clear that, for the $\alpha$ and $\beta$ bands comprising of d$_{xz}$ and d$_{yz}$ orbitals, the band splitting is proportional to $\sin{kx}\sin{ky}$. This explains why the splitting of the $\alpha$ band is smaller than that of the $\beta$ band because the $\alpha$ Fermi surface has a smaller radius.

\section{Superconducting Order Parameters in Considering Interlayer Pairing}

\setlength{\parindent}{2em} Here, we first consider the order parameter of $\beta$ band in the single FeAs layer. The major orbit composition of $\beta$ band is given as:
\begin{equation}
\psi_{\beta}(k)=e^{i\alpha(k)}(\cos\theta\psi_{xz}(k)-\sin\theta\psi_{yz}(k))
\end{equation}
where $\alpha(k)$ is the gauge of the $\beta$ band; $\psi_{xz}$ and $\psi_{yz}$ represent wave functions of the $d_{xz}$ and $d_{yz}$ orbitals; $\theta=arctan\frac{k_x}{k_y}$.
Obviously, the absolute value of the intraband order parameter in $\beta$ band is
\begin{equation}
\Delta_{\beta}(k)=(\cos\theta)^2\Delta_{xz}(k)+(\sin\theta)^2\Delta_{yz}(k)-\sin\theta\cos\theta\Delta_{xz,yz}(k)
\end{equation}
where $\Delta_{xz}$ and $\Delta_{yz}$ are the order parameters of the $d_{xz}$ and $d_{yz}$ orbitals, respectively; $\Delta_{xz,yz}(k)$ is the order parameter of the interorbital pairs, given as:
\begin{equation}
\Delta_{xz,yz}(k)=<\psi_{xz}(k)\psi_{yz}(-k)+\psi_{xz}(-k)\psi_{yz}(k)>
\end{equation}

\setlength{\parindent}{2em} Now, we consider the intraorbital order parameter of $\beta1$ band and $\beta2$ band in K12442 system. In a familiar way, the main orbital composition of $\beta1$ and $\beta2$ bands can be calculated based on the above tight binding model:
\begin{equation}
\begin{split}
\psi_{\beta1}(k)=\frac{\sqrt{2}}{2}e^{i\alpha_1(k)}(\cos\theta\psi_{1,xz}(k)-\sin\theta\psi_{1,yz}(k)+\cos\theta\psi_{2,xz}(k)-\sin\theta\psi_{2,yz}(k))\\
\psi_{\beta2}(k)=\frac{\sqrt{2}}{2}e^{i\alpha_2(k)}(\cos\theta\psi_{1,xz}(k)-\sin\theta\psi_{1,yz}(k)-\cos\theta\psi_{2,xz}(k)+\sin\theta\psi_{2,yz}(k))
\end{split}
\end{equation}
where $\alpha_1(k)$ and $\alpha_2(k)$ are the gauge of $\beta1$ and $\beta2$ bands, respectively; $\psi_{1,xz}(k)$ represents the wave function of $d_{xz}$ in layer 1, the rest $\psi$s are in a familiar way; $\theta=arctan\frac{k_x}{k_y}$.
Based on the above equations, we obtain the absolute value of the order parameter in $\beta1$ and $\beta2$ bands:
\begin{equation}
\begin{split}
\Delta_{\beta1}(k)=\Delta_{\beta}(k)+\Delta_{interlayer}(k)\\
\Delta_{\beta2}(k)=\Delta_{\beta}(k)-\Delta_{interlayer}(k)
\end{split}
\end{equation}
where the $\Delta_{interlayer}(k)$ is the order parameter of the superconducting pairs between the two different layers:
\begin{equation}
\begin{split}
\Delta_{interlayer}(k)=&(\cos\theta)^2<\psi_{1,xz}(k)\psi_{2,xz}(-k)+\psi_{1,xz}(-k)\psi_{2,xz}(k)>+\\&(\sin\theta)^2<\psi_{1,yz}(k)\psi_{2,yz}(-k)+\psi_{1,yz}(-k)\psi_{2,yz}(k)>-\\&\sin\theta\cos\theta<\psi_{1,xz}(k)\psi_{2,yz}(-k)+\psi_{1,xz}(-k)\psi_{2,yz}(k)+\\&\psi_{2,xz}(k)\psi_{1,yz}(-k)+\psi_{2,xz}(-k)\psi_{1,yz}(k)>
\end{split}
\end{equation}